\documentclass[12pt,journal]{article}
\usepackage{graphicx,amssymb,amsmath,amsthm,cite,cases}
\usepackage{color,multirow}
\usepackage{mathrsfs}

\def\Join{\hat{\operatorname{J}}}

\def\vp{\mathbf{p}}

\def\be{\begin{equation}}
\def\ee{\end{equation}}
\def\ben{\begin{eqnarray}}
\def\een{\end{eqnarray}}
\def\Sti{st_{\rm{ind}}}
\def\Stc{st_{\rm{cf}}}
\def\Sts{st_{\rm{sg}}}

\def\D{\mathcal{D}}
\def\R{\mathbb{R}}
\def\N{\mathbb{N}}

\def\crm{\mathrm{CR_{\mathrm{M}}}}
\def\crd{\mathrm{CR_{\mathrm{D}}}}
\def\snr{\mathrm{snr}}
\def\SNR{\mathrm{SNR}}
\def\gqm{\SNR_\mathrm{M}}
\def\gqd{\SNR_\mathrm{D}}
\def\lqm{\overline{\snr}_\mathrm{M}}
\def\lqd{\overline{\snr}_\mathrm{D}}
\def\std{\mathrm{std}}
\def\sdm{\std_\mathrm{M}}
\def\sdd{\std_\mathrm{D}}

\def\til{\tilde}

\def\vd{\mathbf{d}}
\def\vs{\mathbf{s}}
\def\vst{\mathbf{\til{s}}}
\def\st{\til{s}}
\def\ellt{\til{\ell}}

\def\vc{\mathbf{c}}
\def\vct{\mathbf{\til{c}}}

\def\ctq{\til{c}^\Delta}
\def\ct{\til{c}}
\def\vf{\mathbf{f}}
\def\vfr{\mathbf{f}^{\rm{r}}}
\def\vfk{\mathbf{f}^{\kq}}

\def\vB{\mathbf{b}}

\def\vW{\mathbf{w}}
\def\vWt{\til{\mathbf{w}}}

\def\vr{\mathbf{r}}
\def\vd{\mathbf{d}}

\def\kq{k_q}
\def\kqt{\til{k}_q}

\def\Nq{N_b}

\newcommand{\la}{\langle}
\newcommand{\ra}{\rangle}

\title{A dedicated codec for compression of 
Gravitational Waves Sound}

\author{Laura Rebollo-Neira\\
Mathematics Department\\ 
Aston University\\ B4 7ET Birmingham, UK}
\date{}
\begin{document}

\maketitle


\begin{abstract}
A dedicated codec for compression of gravitational 
waves sound with high quality recovery is proposed. 
The performance is tested on the available set of gravitational sound signals that has been theoretically generated at the Massachusetts Institute of Technology (MIT).  
The approach is based on a model for data reduction rendering high quality approximation of the signals.  The reduction of dimensionality is achieved by selecting elementary components from a redundant set called a dictionary.  Comparisons with the compression standard MP3 demonstrate the merit of the dedicated technique for compressing this type of sound. 
\end{abstract}
\section{Introduction}
 After the celebrated first detection of a gravitational 
 wave (GW) on Sept 2015 \cite{LV115} five more detections 
 \cite{LV216,LV317,LV417,LV517,LV617} have confirmed Einstein's theory 
 of general relativity. Predictions from the Laser 
 Interferometer Gravitational-Wave Observatory (LIGO)
 and Virgo Scientific Collaboration assert that weekly 
 or even more often detections can be expected in the 
 near future. Scientists envisage  the prospect 
 as the beginning of a new era in astronomy, whereby 
 the very early Universe will be studied by the 
 sound that was made in its formation.  
 Theoretically, GWs are 
 modelled and generated using techniques of numerical 
 relativity \cite{Pre0,CLM06,BCC06}, from which audio 
 representation  can be developed. 
 In particular, the group of 
 Prof. Hughes at MIT has produced and made 
   available the gravitational waves sound (GWS) 
 signals which have been used in this study. 
 The numerically simulated 
 signals are supported by a number of publications 
\cite{Hug00, Hug01, GHK02, HDF05, DH06} 
and nicely presented on a website \cite{MITData}.
This work uses those signals to demonstrate a dedicated 
scheme for the compression of GWS with high quality 
recovery. 

The  proposal falls within the
usual transform coding scheme. It consists of three
main steps:
1) Transformation of the sound signal.
2) Quantization  of the transformed data.
3) Bit-stream entropy coding.
However, it differs from the traditional
compression techniques from the beginning.  
Instead of considering an orthogonal transformation,
the first step is realized by approximating
the signal using a redundant dictionary from where 
the elementary components for representing the signal are 
selected through a greedy pursuit strategy.  
This strategy gives rise to what is known as sparse 
 representation of a signal. In the area of audio processing 
 a number of different tasks have been shown to benefit by 
the sparsity of a representation. The relevant literature is
extensive. As a sample we refer to
\cite{RRD08, LPL09, PBD10, NHS12, HJK14, YSY14,VML14,PBD16,HLN16, RLA16, RNS17}. 
Here we focus on compression of GWS signals,  
 which do require a dedicated framework for their sparse 
representation. In particular, the dictionary we 
use for the proposed codec was introduced in a recent publication as 
potentially good to achieve sparse representation
 of GWS by partitioning the signal 
\cite{RNP17}. Now it stands as a crucial component 
of a codec for high quality point-wise 
recovery of the compressed signal.

Since the GWS under consideration is in the range of 
 human 
hearing, it is interesting to realize comparisons 
 with the popular compression standard MP3. 
The results demonstrate  
a significant improvement in compression performance for 
the equivalent quality of the recovered signal. 
 More precisely, 
the signal is required to yield a
Signal to Noise
Ratio (SNR) competitive with MP3 outcomes
with respect to both,
the whole signal and the elements of the signal
partition. 

The benefits of the proposed dedicated 
format for encoding GWS  go beyond
the compression performance.
Certainly, the underlying representation of the 
data generates a reduced set which contains  
 information about the elementary components of the 
signal. Hence, in addition to recovering the signal from 
a compressed file with high quality, one can also recover 
its reduced representation in terms of elementary components.
 Since the reduced representation  
 gives rise to an accurate approximation of 
the signal, it can be of assistance to signal 
processing techniques relying on a reduction of 
dimensionality as a first step of further processing. 

The main goals of the numerical tests of Section \ref{NR}
are:
\begin{itemize}
\item To produce strong evidence that 
high quality approximation of two differently generated 
types of GWS can be achieved as 
 a superposition of elementary components 
of different nature. 
A component of elements generated by a discrete 
version of trigonometric functions, 
 and a component consisting of pulses of small support.

\item To demonstrate that the above described decomposition 
can be stored in a file which is significantly smaller 
than that obtained with the  compression 
standard MP3. 
\end{itemize}
The goals are achieved by testing the method on 
the available  audio representation of GWs which 
are grouped, according to their theoretical modeling,
as follows \cite{MITData}:
\begin{itemize}
\item
Extreme mass ratio inspiral (EMRI).  
Gravitational waves produced when a relatively light compact 
object orbits around a much heavier black hole and gradually 
decays.
\item
Binaries. Emitted during the 
merger of two bodies of roughly the same mass. 
The first direct detection
of a GW belongs to this category.
\end{itemize}
\section{Method}
\label{Met}
The principal constituent of the proposed codec is the
mathematical model of the sound signal. As already mentioned,  the model is realized by selecting elementary components
 of a dedicated  dictionary containing atoms of
 different nature. Thus, the method for
selecting the suitable atoms for the approximation
of a given signal is also relevant to the modelling.
 For this task we use the Optimized Orthogonal Matching Pursuit (OOMP) method \cite{RNL02}, which is step wise optimal
in the sense of minimizing the norm of the approximation
error at each iteration step. 

Throughout the description of the methods $\R$
and $\N$ stand for
the sets of real and natural numbers, respectively.
Boldface letters are used
to indicate Euclidean vectors  and
their corresponding components are represented using
standard  mathematical fonts,
e.g., $\vf \in \R^N,\, N \in \N$ is a vector of components
$f(i),\, i=1,\ldots,N$.  The
 inner product operation is indicated
as $\la \cdot, \cdot \ra$.

A partition of a signal
$\vf \in \R^{N}$ is realized by
 a set of disjoint pieces $\vf_q \in \R^{\Nq},\, 
q=1,\ldots,Q$, which for simplicity are assumed to
be all of the same size and such that $Q \Nq=N$,
i.e., it holds that
$\vf = \Join_{q=1}^{Q} \vf_{q}$, where the
concatenation operation $\Join$ is
defined as follows: $\vf$ is a vector
in $\R^{Q \Nq}$ having components
$f(i)=f_{q}(i-(q-1)\Nq),\, i=(q-1)\Nq+1,\ldots,q\Nq,\,q=1,\ldots,Q$. Hereinafter each element of the signal partition 
$\vf_q$ will be refereed  to as a `block'.
\subsection{OOMP approximation}
Given a signal $\vf $
partitioned into $Q$ blocks $\vf_{q} \in \R^{\Nq},\,q=1,\ldots,Q$,
, the $\kq$-term approximation
of each block is modelled by the superposition
\be
\label{atoq}
\vfk_{q}= \sum_{n=1}^{\kq}
c^q(n) \vd_{\ell^{q}_n},
\quad  q=1,\ldots,Q.
\ee
The elements  $\vd_{\ell^{q}_n},\,n=1,\ldots,\kq$ in
\eqref{atoq}, called 'atoms' 
are selected here from a dedicated dictionary
$\D =\{\vd_n \in \R^{\Nq},\, \| \vd_n\|=1\}_{n=1}^{M}$ through the
  OOMP approach \cite{RNL02} which, for each block $q$ operates as
follows.
The algorithm is initialized by setting:
$\vr_q^0=\vf_q$, $ \vf_q^0=0$, $\Gamma_q= \emptyset$
and $\kq=0$. The
 first atom for
 the atomic decomposition of the $q$-th block
 is selected as the one corresponding to the index
$\ell_{1}^q$ such that
\be
\ell_{1}^q=\operatorname*{arg\,max}_{n=1,\ldots,M}
 \left |\la \vd_n,\vr_{q}^{\kq}\ra \right|^2.
\ee
This first atom is used to
assign $\vW_1^q=\vB_1^{1,q}=\vd_{\ell_{1}^q}$,
calculate $\vr_{q}^{1}= \vf_q - \vd_{\ell_{1}^q} 
\la \vd_{\ell_{1}^q}, \vf_q\ra$ and  iterate
 as prescribed below.
\begin{itemize}
\item[1)] Upgrade the set $\Gamma_{q} \leftarrow  \Gamma_{q} \cup 
\ell_{\kq+1}$, increase $\kq\leftarrow \kq +1$, and
 select the index of a new atom for the approximation as
\be
\label{oomp}
\ell_{\kq+1}^{q}=\operatorname*{arg\,max}_{\substack{n=1,\ldots,M\\ n\notin \Gamma_{q}}}
 \frac{|\la \vd_n,\vr_{q}^{\kq}
\ra|^2}{1 - \sum_{i=1}^{\kq}
|\la \vd_n ,\vWt_i^{q}\ra|^2},
 \quad  \text{with} \quad \vWt_i^{q}= \frac{\vW_i^{q}}{\|\vW_i^{q}\|}.
\ee
\item[2)]
Compute the corresponding new vector $\vW_{\kq+1}^{q}$ as
\be
\begin{split}
\label{GS}
\vW_{\kq+1}^{q}= \vd_{\ell_{\kq+1}}^{q} - \sum_{n=1}^{\kq} \frac{\vW_n^q}
{\|\vW_n^q\|^2} \la \vW_n^q, \vd_{\ell_{\kq+1}}^{q}\ra.
\end{split}
\ee
including, for numerical accuracy,  the
re-orthogonalizing step:
\be
\label{RGS}
\vW_{\kq+1}^{q} \leftarrow \vW_{\kq+1}^{q}- \sum_{n=1}^{\kq} \frac{\vW_{n}^q}{\|\vW_n^q\|^2}
\la \vW_{n}^q , \vW_{\kq+1}^{q}\ra.
\ee
\item[3)]
Upgrade vectors
$\vB_n^{\kq,q}$ as
\be
\begin{split}
\label{BW}
\vB_{n}^{{\kq}+1,q}&= \vB_{n}^{{\kq},q} - \vB_{\kq+1}^{{\kq}+1,q}\la \vd_{\ell_{{\kq}+1}}^{q}, \vB_{n}^{\kq+1,q}\ra,\quad n=1,\ldots,\kq,\\
\vB_{\kq+1}^{\kq+1,q}&= \frac{\vW_{\kq+1}^q}{\| \vW_{\kq+1}^q\|^2}.
\end{split}
\ee
\item[4)]
Calculate
\ben
\vr_{q}^{\kq+1} &=& \vr_{q}^{\kq} - \la \vW_{\kq+1}^{q}, \vf_{q} \ra  \frac{\vW_{\kq+1}^{q}}{\| \vW_{\kq+1}^{q}\|^2}.
\een
\item[5)]If for  a given
 $\rho$ the condition
 $\|\vr_{q}^{\kq+1}\| < \rho$ has been met
 compute the coefficients
$c^{k_q}(n) = \la \vB_n^{\kq}, \vf_q \ra,\, n=1,\ldots, \kq$.
 Otherwise repeat steps 1) - 5).
\end{itemize}
{\bf{Remark 1:}} For each value of $q$
the criterion \eqref{oomp} in the method above gives the 
index minimizing the local
residual norm $\|\vf_q -\vf_q^{\kq}\|$  with respect to 
the newly selected atom. Moreover, the 
 approximation of each block $q$ is completed at once
totally independent of the other blocks. In previous 
publications \cite{LRN16,RNP17} we have shown the benefit 
in ranking the blocks for their 
 sequential stepwise approximation to minimize the total residual error $\|\vf - \vf^K\|$, with $K= \sum_{q=1}^Q 
\kq$. Such a strategy is called
 Optimized Hierarchical Block Wise 
OOMP (OHBW-OOMP).  However, 
for the particular application to 
compression, the quantization procedure plays also a 
role in the approximation by mapping small coefficients 
to zero and quantizing the others. Consequently, 
for the set of signals considered in this work 
 the compression results rendered by both  methods 
 are practically equivalent.
\subsection{The Dictionary}
\label{dict}
The dictionary used for the approximation consists of
 two sub-dictionaries  of different nature.
One of them is a trigonometric dictionary $\mathcal{D}_T$,
which is the union of the dictionaries $\mathcal{D}_{C}$ and
$\mathcal{D}_{S}$ given below.
\ben
\mathcal{D}_{C}^x&=&\{w_c(n)
\cos{\frac{{\pi(2i-1)(n-1)}}{2M}},i=1,\ldots,\Nq\}_{n=1}^{M}\nonumber\\
\mathcal{D}_{S}^x&=&\{w_s(n)\sin{\frac{{\pi(2i-1)(n)}}{2M}},i=1,\ldots,\Nq\}_{n=1}^{M},\nonumber
\een
where $w_c(n)$ and $w_s(n),\, n=1,\ldots,M$ are
normalization factors. In the numerical simulations we
have considered $M=2 \Nq$ and $\Nq=2048.$

\begin{figure}[!ht]
\begin{center}
\includegraphics[width=9cm]{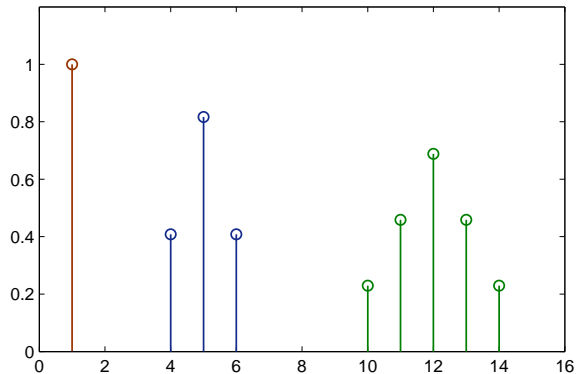}
\caption{\small{Prototype atoms $\vp_1, \vp_2$ and
$\vp_3$, which generate the
dictionaries $\mathcal{D}_{P_1}$, $\mathcal{D}_{P_2}$
and $\mathcal{D}_{P3}$ by sequential
translations of one point.}}
\label{pato}
\end{center}
\end{figure}

The other sub-dictionary is constructed
by translation of the prototype atoms, $\vp_1, \vp_2$ and
$\vp_3$ in  Fig.~\ref{pato}.

Denoting  by
$\mathcal{D}_{P_1}$, $\mathcal{D}_{P_2}$
and $\mathcal{D}_{P_3}$ the
dictionaries  arising  by translations of the atoms
$\vp_1$, $\vp_2$,  and $\vp_3$, respectively,
the dictionary $\mathcal{D}_{P}$ is
built as
$\mathcal{D}_{P}= \mathcal{D}_{P_1} \cup \mathcal{D}_{P_2}
\cup \mathcal{D}_{P_3}$.
The whole dictionary is then built as
$\mathcal{D} = \mathcal{D}_{T} \cup 
\mathcal{D}_{P}$, with
$\mathcal{D}_{T}= \mathcal{D}_{C} \cup \mathcal{D}_{S}$.
This dictionary was previously introduced
as potentially suitable for approximation
of GWS \cite{RNP17}.
In this study it stands out as
 being  essential for achieving high quality
approximation of most of the signals which have been
compressed with the proposed codec.
\subsection{Coding Strategy}
\label{cod}
Previously to entropy encoding the coefficients resulting
from approximating a signal by partitioning, the
real numbers need to be converted into integers. This
operation is known as quantization. For all the
numerical cases  we have adopted a simple
uniform quantization technique.
The absolute value coefficients
$|c_{q}(n)|,\,n=1\ldots,\kq,\,q=1,\ldots,Q,
$ are
converted to integers as follows:
\be
\label{uniq}
c^\Delta_{q}(n)= \lfloor \frac{|c_{q}(n)|}{\Delta} +\frac{1}{2} \rfloor,
\ee
where $\lfloor x \rfloor$ indicates the largest
integer number
smaller or equal to $x$ and  $\Delta$ is the quantization
parameter.
The signs of the coefficients, represented as
$\vs_{q},\,q=1,\ldots,Q$,
are encoded separately using a binary alphabet.

For creating the stream of numbers to encode
 the signal model we proceed as in
a recent work \cite{RNS17}.
The indices of the atoms in the
 atomic
decompositions of each block $\vf_q$
 are first sorted in ascending
order
$\ell_{i}^q \rightarrow \til{\ell}_i^q,\,i=1,\ldots,\kq$,
which guarantees that, for each $q$ value,
$\til{\ell}_i^q < \til{\ell}_{i+1}^q,\,i=1,\ldots,\kq-1$.
This order of the indices induces
an order in the unsigned coefficients,
$\vc^\Delta_{q} \rightarrow \vct^\Delta_{q}$ and
in the corresponding signs $\vs_{q} \rightarrow
\vst_{q}$.  The ordered indices are
stored as smaller positive numbers by taking differences
between two consecutive values.
By defining $\delta^q_i=\ellt^q_i- \ellt^q_{i-1},\,i=2,\ldots,\kq$ the follow string stores the indices for block
$q$ with unique recovery
${\ellt^q_1, \delta^q_2, 
\ldots, \delta^q_{\kq}}$.
The number `0' is then used to separate the string
corresponding to different blocks and entropy code
 a long string, $\Sti$, which is built as
\be
\label{sti}
\Sti=[{\ellt_1}^1,\ldots,\delta^1_{k_1},0,
{\ellt_1}^2,\ldots,\delta^2_{k_2},0, \cdots,
{\ellt_1}^{k_Q},\ldots, \delta^Q_{k_Q}].
\ee
The corresponding quantized magnitude of the coefficients
are concatenated in the strings $\Stc$
 as follows:
\be
\label{stc}
\Stc
= [\ctq_{1}(1), \ldots, \ctq_{1}(k_1), \cdots,
\ctq_{{k_Q}}(1), \ldots, \ctq_{{k_Q}}(k_Q)].
\ee
Using `0' to store a positive sign and `1' to store
negative one, the signs  are placed
in the string, $\Sts$  as
\be
\label{sts}
\Sts=[\st_{1}(1), \ldots, \st_{1}(k_1), \cdots,
\st_{{k_Q}}(1), \ldots, \st_{{k_Q}}(k_Q)].
\ee
The next encoding$/$decoding scheme
summarizes the above described procedure.\\\\
{\bf{Encoding}}
\begin{itemize}
\item
Given a  partition
$\vf_q \in \R^{\Nq},\, q=1,\ldots,Q$
of a signal,
approximate each block $\vf_q$ by the atomic
decompositions \eqref{atoq}.
\item
Quantize, as in \eqref{uniq},
the absolute value coefficients 
 to obtain
$c_{q}^{\Delta} (n),\,n=1,\ldots,\kq,\, 
q=1,\ldots,Q$.
\item
For each $q$, sort the indices
$\ell_1^q,\ldots,\ell_{\kq}$ in ascending
order to have a new order $\ellt_1^q,\ldots, \ellt_{\kq}$
and the re-ordered sets $\st_{q}(1), \ldots, \st_{q}(\kq)$,
and $\ct^{\Delta}_{q}(1),\ldots,\ct^{\Delta}_{q}(\kq)$,
to create the strings:
$\Sti$, as in \eqref{sti}, and
$\Stc$, and $\Sts$ as in \eqref{stc} and
\eqref{sts}, respectively. All these strings are
 encoded, separately, using adaptive arithmetic coding.
\end{itemize}
{\bf{Decoding}}
\begin{itemize}
\item
Reverse the arithmetic coding to recover strings
 $\Sti, \Stc, \Sts$.
\item
Invert the quantization step as
\be
|\til{c}_{q}^{\mathrm{r}}(n)|=
\Delta {\til{c}}_{q}^\Delta(n).
\ee
\item
Recover the approximated partition
through the linear combination
\be
\vf_{q}^{{\mathrm{r}},\kq}
 = \sum_{n=1}^{\kq} \til{s}_{q}(n)|{\til{c}}_{q}^{\mathrm{r}}(n)| \vd_{\til{\ell}_{n}^q}.
\ee
\item
Assemble the recovered signal  as
\be
\vf^{\mathrm{r}}=\Join_{q=1}^Q \vf_{q}^{{\mathrm{r}},\kq}.
\ee
\end{itemize}

\section{Results}
\label{NR}
Given the approximation $\vfr$ of a signal $\vf$, the quality of such an approximation is assessed by the SNR 
 calculated as
\be
\text{SNR}=10 \log_{10} \frac{\| \vf\|^2}{\|\vf - \vfr\|^2},\quad
\ee
where $\| \cdot\|$ indicates the 2-norm.  

The local SNR with respect to every block in the
partition, which we indicate as
$\text{snr}(q),\,q=1,\ldots Q$,
is calculated as
\be
\text{snr}(q)=10 \log_{10} \frac{\| \vf_q\|^2}{\|\vf_q - 
\vf_q^{{\mathrm{r}},\kq}\|^2},
\ee
where $\vf_q^{{\mathrm{r}},\kq}$ is the approximation 
of the block $\vf_q$.
Both, the mean value ($\overline{\text{snr}}$)
 and standard deviation (std) of these 
 local quantities are relevant to comparison of
 point-wise quality recovery. Accordingly, we define
\be
\overline{\text{snr}}=\frac{1}{Q}\sum_{q=1}^Q \text{snr}(q),\quad \text{and} \quad 
\text{std}= \sqrt{\frac{1}{Q-1}\sum_{q=1}^Q (\text{snr}(q)
- \overline{\text{snr}})^2}.
\ee
In order to use the SNR and $\overline{\text{snr}}$ as measures of quality for comparison, the MP3 signal has to
 be optimized in relation to those quantities.
This is carried
out by the following operations\cite{RNS17}.
\begin{itemize}
\item{Shifting}: Since  MP3 introduces a shift
 with respect to the original signal, to compute the SNR 
 that shift should be reversed. 
 Denoting by ${\vf}_{\mathrm{M}}$ the
numerical signal
retrieved from the MP3 file, 
the optimal time shift $\hat{\tau}$
 is determined to be the time shift maximizing the cross-correlation
with the original signal, i.e.
\be
\hat{\tau} = \operatorname*{arg\,max}_{\tau=-N/2,\ldots,N/2}\, \sum_{n=1}^{N}  f(n)   f_{\mathrm{M}}(n+\tau).
\ee
\item{Scaling}: In order to neutralize the effect of any
multiplicative and/or additive constant which could affect
the SNR value, we allow for such two constants and adjust
them to maximize the SNR as follows: 
 Denoting by $\hat{\vf}_{\mathrm{M}}$ the MP3 signal after 
the shifting operation,  we consider the linear form
$a \hat{\vf}_{\mathrm{M}} + b$ and fix the values of $a$ and
$b$  for which $\| \vf - (a \hat{\vf}_{\mathrm{M}} + b)\|^2$ takes the minimum value, i.e.
\ben
a&=&\frac{\la \vf, \vf_{\mathrm{M}}\ra - (\frac{1}{N}\sum_{i=1}^N \hat{f}_{\mathrm{M}}(i))(\frac{1}{N}\sum_{j=1}^N f(j))}{\|\hat{\vf}_{\mathrm{M}}\|^2 - \frac{1}{N} (\sum_{i=1}^N f_{\mathrm{M}}(i))^2} \nonumber\\
b&=& \frac{1}{N}\sum_{i=1}^N f(i)- a \frac{1}{N}\sum_{i=1}^N \hat{f}_{\mathrm{M}}(i). \nonumber
\een
\end{itemize}
While the additive constant $b$ is not relevant, 
 the scaling constant $a$ produces an important 
correction which significantly increases the SNR.

The compression power is determined by 
 the compression ratio 
(CR) defined as follows
\be
\text{CR}=\frac{{\text{Size of the file with the signal}}}{{
\text{Size of the compressed file}}}.
\ee
In all the numerical examples the files of 
the signals are given in WAV format. 
The compression is realized on a single audio channel.
 The mean value of all the signals is practically zero. 

As discussed below, after the shifting and scaling operation 
for low values of CR (e.g. CR=2 and CR=4) MP3  recovers 
signals of high quality with respect to both measures, the 
SNR and the $\overline{\text{snr}}$. The proposed 
codec, henceforth to be refereed to as `dictionary codec' (DC), is set to produce an approximation of the 
original signal matching either the 
 SNR  or the  
$\overline{\text{snr}}$ value achieved by the MP3 signal 
(whatever quantity is the largest one). As seen in the tables below, 
by matching the largest quantity of either SNR, or $\overline{\text{snr}}$, the other measure is guaranteed to be 
 larger than the value attained by the  MP3 signal.

\subsection{Numerical Case I}
As a first numerical example we consider 
the audio representation of  a 
detected GW, the chirp {\tt{gw151226}} \cite{LigoData}.  
This is a short signal, 
it consists of $N= 65536$ points with sampling 
frequency of 41 kHz. For CR=2 the point-wise quality 
of the signal recovered from the MP3 file is 
 excellent: SNR=74.6 dB and $\overline{\text{snr}}=70.2$~dB.  These values should be appreciated by taking into account 
that  
SNR=74.6 dB corresponds to a mean square error between the 
approximation and the signal of order $10^{-9}$.
Since in this case the SNR is greater than 
the $\overline{\text{snr}}$ the DC is set 
to produce the same value 
of SNR. With this restriction the resulting 
 $\overline{\text{snr}}$ is 71.7dB, i.e. larger than 
the value yielded by MP3. The huge difference between the 
two encoding procedures is the CR. Denoting by  $\crm$ the 
CR with respect to the MP3 file and $\crd$ that of the 
DC file, for SNR=74.6 dB one has $\crm=2$ and 
$\crd=23!$ While for larger values of $\crm$ the quality
 of the recovered signal decreases, up to $\crm=10$ 
 the quality of the MP3 signal is still very good. 
 Table~\ref{TABLE1} displays the values of SNR and 
$\overline{\text{snr}}$, as well as 
the corresponding CRs produced by both formats. These 
are differentiated by the notation $\gqm$ and $\lqm$, 
used to indicate the values produced by the MP3 format, 
and $\gqd$ and $\lqd$, used to indicate the values produced by the DC format.

\begin{table}[h!]
\begin{center}
\begin{tabular}{|c|c|c|c|c|c|r|r||}
\hline
$\gqm$& $\gqd$ &$\lqm$& $\sdm$&$\lqd$& $\sdd$&$\crm$& $\crd$\\ \hline \hline
74.5 dB &  74.5 dB&   70.2 dB&  6.3&  71.7 dB&  5.9&   2.1   &  23.2 \\ \hline
73.8 dB  &  73.8 dB&   69.8 dB&  6.4&  71.1 dB&  5.7&   4.2   &  24.9\\ \hline
71.3 dB &  71.3 dB&   66.9 dB&  6.7&  68.4 dB&  5.8&   8.4   &  61.2\\ \hline
69.4 dB &  69.4 dB&   65.2 dB&  6.2&  66.4 dB&  5.9&   10.4  &  69.3 \\ \hline
\end{tabular}
\caption{Comparison of  $\crm$ and $\crd$,  
 for the sound representation of the detected     
 chirp {\tt{gw151226}}. The first column gives the values of 
 $\gqm$ produced by the MP3 signal recovered from files 
with 
$\crm$ as listed in the 7th column. 
The second column are the identical values of  
 $\gqd$  produced 
 by the signal recovered from the DC files with 
 $\crd$  as listed in 
 the last column.  The 3rd and 5th columns 
are the values of $\lqm$  and $\lqd$, respectively. The 
4th and 6th columns are the corresponding standard deviations.}
\label{TABLE1}
\end{center}
\end{table}

\subsection{Numerical Case II}
In this case the group of GWS has been numerically 
simulated at MIT
\cite{MITData}. The GWs belong to the EMRI category with 
 circular orbit. The signals are organized in two large 
 subgroups, according to the spin of the larger black hole: 
 spin $99.8\%$ of the maximum value and 
 spin $35.94\%$ of the maximum value. Each subgroup 
contains 16 signals, each of which is characterized by 
two angles: the orbital plane and the viewing angle. 
The signals corresponding to spin $99.8\%$ are listed 
in the first column Table~\ref{TABLE2} and 
Table~~\ref{TABLE3}.  The
signals corresponding to spin $35.94\%$ are listed in 
the first column of Table~\ref{TABLE4} and Table~\ref{TABLE5}. 
The frequency of these signals is 8kHz and 
were compressed with MP3
 at the lowest possible rate,  $\crm=2$, 
in the first instance.
 The recovered signals produce the 
values of $\lqm$ and $\gqm$ given in the second and 
sixth columns of Table~\ref{TABLE2}. Since in this case $\lqm > \gqm$
the DC was set to produce $\lqd=\lqm$. 
This guarantees that $\gqd> \gqm$ for all  signals.
The actual values of $\lqm$ vary with the signals, but for 
all of them the recovery is of good quality. 
It corresponds to mean square errors of order of 
$10^{-8}$. As can be observed in the table, 
for all the signals   
the compression performance of the DC format is  clearly 
superior to  MP3. 
Table~\ref{TABLE3} displays the equivalent results but corresponding to 
 a larger compression ratio of MP3 ($\crm=4$). 
\begin{table}[h!]
\begin{center}
\begin{tabular}{|l||r|r|r|r|r|r|c|r||}
\hline
 Signal& $\lqm$& $\sdm$&$ \lqd$& $\sdd$&$ \gqm$&$ \gqd$& $\crm$& $\crd$\\ \hline \hline
E1 &67.9 &2.1 &67.9& 0.2& 65.8&  67.8&  2&  6.8\\ \hline
E2 &65.0 &2.9 &65.0 &0.4& 58.9 & 64.9&  2 & 5.3\\ \hline
E3 &63.7 &3.3 &63.7 &0.5& 54.4 & 63.6&  2 & 5.2\\ \hline
E4 &66.9 &3.4 &66.9 &0.6& 55.2 & 66.7&  2 & 5.6\\ \hline
E5 &63.4 &1.8 &63.4 &0.2& 62.4 & 63.4&  2 & 6.0\\ \hline
E6 &62.2 &2.9 &62.3 &0.2& 58.9 & 62.1&  2 & 5.1\\ \hline
E7 &61.6 &2.9 &61.6 &0.2& 57.9 & 61.6&  2 & 4.9\\ \hline
E8 &62.8 &2.9 &62.8 &0.3& 59.4 & 62.8&  2 & 5.2\\ \hline
E9 &65.7 &2.2 &65.7 &0.3& 64.5 & 65.7&  2 & 6.4\\ \hline
E10&64.1 &2.1 &64.1 &0.2& 62.4 & 64.1&  2 & 5.6\\ \hline
E11&62.7 &1.8 &62.7 &0.2& 62.2 & 62.7&  2 & 5.5\\ \hline
E12&63.7 &2.2 &63.7 &0.2& 63.0 & 63.7&  2 & 5.8\\ \hline
E13&68.5 &4.4 &68.5 &0.4& 64.3 & 68.5&  2 & 6.3\\ \hline
E14&66.8 &2.4 &66.8 &0.4& 65.8 & 66.8&  2 & 5.6\\\hline 
E15&63.7 &2.7 &63.7 &0.6& 62.7 & 63.6&  2 & 5.7\\ \hline
E16&64.5 &2.9 &64.5 &0.4& 62.6 & 63.7&  2 & 5.9\\ \hline \hline
\end{tabular}
\caption{Comparison of $\crm$ and $\crd$ values for GWs within the EMRI category for circular orbit and spin $99.8\%$ of the maximum value. Each signal listed in the first column corresponds to a
particular orbital plane and viewing angle. 
The second column shows the $\lqm$ values produced by
the MP3 signals recovered from a file corresponding 
to $\crm=2$. The sixth column shows the corresponding values 
of $\gqm$. All the other figures in the table
 arise when setting the DC to achieve $\lqd=\lqm$.
}
\label{TABLE2}
\end{center}
\end{table}

\begin{table}[h!]
\begin{center}
\begin{tabular}{|l||r|r|r|r|r|r|c|r||}
\hline
Signal&$\lqm$& $\sdm$&$ \lqd$&$ \sdd$& $\gqm$&$\gqd$&$\crm$&$ \crd$ \\ \hline
E1 & 60.3 & 2.4&  60.3&  0.3&  58.2&   60.2&   4&   9.7\\ \hline
E2 & 55.5 & 4.2&  55.5&  0.4&  47.9&   55.4&   4&   7.9\\ \hline
E3 & 54.1 & 4.9&  54.1&  0.4&  45.5&   54.0&   4&   7.6\\ \hline
E4 & 57.5 & 4.0&  57.5&  0.6&  51.5&   57.4&   4&   8.3\\ \hline
E5 & 54.9 & 2.9&  54.9&  0.2&  52.4&   54.8&   4&   8.8\\ \hline
E6 & 52.4 & 4.2&  52.4&  0.2&  44.9&   52.3&   4&   7.6\\ \hline
E7 & 52.2 & 5.0&  52.2&  0.2&  39.7&   52.1&   4&   7.1\\ \hline
E8 & 53.6 & 3.9&  53.6&  0.3&  43.0&   53.6&   4&   7.5\\ \hline
E9 & 56.5 & 2.7&  56.5&  0.2&  55.3&   56.4&   4&   9.6\\ \hline
E10& 54.7 & 1.8&  54.7&  0.2&  53.8&   54.6&   4&   8.2\\ \hline
E11& 53.2 & 2.4&  53.2&  0.2&  51.9&   53.2&   4&   8.1\\ \hline
E12& 54.4 & 2.8&  54.4&  0.2&  52.0&   54.4&   4&   8.4\\ \hline
E13& 59.4 & 3.4&  59.4&  0.4&  57.3&   59.4&   4&   8.8\\ \hline
E14& 57.4 & 2.2&  57.3&  0.3&  56.7&   57.4&   4&   7.7\\ \hline
E15& 55.5 & 2.2&  55.5&  0.3&  51.2&   55.5&   4&   7.7\\ \hline
E16& 55.0 & 2.6&  55.0&  0.5&  51.2&   55.0&   4&   8.4\\ \hline \hline
\end{tabular}
\caption{Same  description as in Table~2 but for $\crm=4$.}
\label{TABLE3}
\end{center}
\end{table}

\vspace{4cm}

Table~\ref{TABLE4} and Table~\ref{TABLE5} have equivalent description as Table~\ref{TABLE2}
and \ref{TABLE3}, respectively, but the signals belong to  the
EMRI group with spin $35.94\%$ of the maximum value. 
\begin{table}[h!]
\begin{center}
\begin{tabular}{|l||r|r|r|r|r|r|c|r||}
\hline
 Signal& $\lqm$& $\sdm$&$ \lqd$& $\sdd$&$ \gqm$&$ \gqd$& $\crm$& $\crd$\\ \hline \hline

S1&  71.8 &  3.2& 71.8 & 0.6 & 68.8&   71.7&  2&   10.2\\ \hline
S2&  70.2 & 2.4&  70.2 & 0.7 & 67.9&   70.1&  2&   6.6\\ \hline
S3&  69.7 &2.2 &  69.7  &0.5 & 67.3&   69.6&  2&   6.4\\\hline
S4&  69.5 & 2.4&  69.5 & 0.5 & 67.3&   69.4&  2&   6.4\\ \hline
S5&  70.8 &  2.3&  70.8&  0.5&  68.9&   70.8&  2&   7.7\\ \hline
S6&  68.8 &  1.9&  68.8&  0.4&  67.9&   68.7&  2&   5.5\\ \hline
S7&  65.0 &  3.3&  65.0&  0.4&  61.1&   65.0&  2&   5.4\\ \hline
S8&  64.8 &  1.9&  64.8&  0.4&  63.9&   64.7&  2&   5.8\\ \hline
S9&  69.9 &  2.5&  69.9&  0.3&  68.2&   69.9&  2&   5.2\\ \hline
S10& 67.1 &  1.7&  67.1&  0.4&  66.6&   67.2&  2&   4.6\\ \hline
S11& 63.4 &  2.3&  63.4& 0.5 & 62.3 &  63.5 & 2 &  4.5\\ \hline
S12& 61.7 &  2.0&  61.7&  0.2&  60.9&   61.7&  2&   4.5\\ \hline
S13& 66.0 &  3.2&  66.0&  0.6&  62.8&   66.1&  2&   4.9\\ \hline
S14& 63.0 &  7.2&  63.0&  1.6&  61.6&   64.4&  2&   4.9\\ \hline
S15& 59.0 &  8.4&  59.0&  1.3&  59.5&   60.2&  2&   4.9\\ \hline
S16& 51.8 &  2.0&  51.8&  0.3&  51.0&   51.9&  2&   4.7\\ \hline
\end{tabular}
\caption{Same  description as for Table~\ref{TABLE2} 
but the signals belong to  the
EMRI group with spin $35.94\%$ of the maximum value.}
\label{TABLE4}
\end{center}
\end{table}

\vspace{2cm}

\begin{table}[h]
\begin{center}
\begin{tabular}{|l||r|r|r|r|r|r|c|r||}
\hline
Signal&$\lqm$& $\sdm$&$ \lqd$&$ \sdd$& $\gqm$&$\gqd$&$\crm$&$ \crd$ \\ \hline
S1& 66.3  & 3.2&  66.3&  0.5&  61.9&   66.4&  4&   13.1\\ \hline
S2& 62.9  & 2.7&  62.9&  0.5&  58.7&   62.8&  4&   9.2\\ \hline
S3&  61.4 &  2.9&  61.4&  0.5&  57.5&   61.2&  4&   9.3\\ \hline
S4&  61.6 &  2.4&  61.6&  0.5&  58.6&   61.5&  4&   9.0\\ \hline
S5&  64.1 &  2.9&  64.1&  0.5&  61.3&   64.0&  4&   10.3\\ \hline
S6&  61.1 &  2.1&  61.1&  0.4&  59.2&   61.0&  4&   7.9\\ \hline
S7&  57.3 &  2.5&  57.3&  0.4&  54.8&   57.3&  4&   7.8\\ \hline
S8&  56.8 &  2.5&  56.8&  0.4&  54.5&   56.8&  4&   8.3\\ \hline
S9&  61.3 &  2.6&  61.3&  0.3&  59.9&   61.2&  4&   8.0\\ \hline
S10& 58.1 &  2.2&  58.1&  0.4&  56.7&   58.1&  4&   7.2\\ \hline
S11& 54.2 &  2.6&  54.2&  0.5&  52.9&   54.3&  4&   7.1\\ \hline
S12& 53.8 &  2.0&  53.8&  0.2&  53.0&   53.8&  4&   6.8\\ \hline
S13& 56.8 &  2.9&  56.8&  0.6&  54.4&   56.8&  4&   7.7\\ \hline
S14& 55.1 &  7.4&  55.1&  1.2&  54.0&   56.1&  4&   7.0\\ \hline
S15& 51.0 &  7.5&  51.0&  1.2&  51.7&   52.0&  4&   7.0\\ \hline
S16& 44.7 &  2.7&  44.7&  0.7&  43.1&   44.6&  4&   6.6\\ \hline \hline
\end{tabular}
\caption{Same as the description of Table~\ref{TABLE4} but for $\crm=4$.}
\label{TABLE5}
\end{center}
\end{table}

\subsection{Numerical Case III}
This group of signals has also been simulated at MIT. All the signals belong to the Binary category, for circular 
systems, and are 
 differentiated by the mass of the larger body.  
The first two signals listed in Tables~\ref{TABLE6} and 
\ref{TABLE7} correspond to bodies of roughly the same 
mass. The first signal (B1) is generated by 
binary neutron stars each of 1.5 solar masses. The second
signal (B2) is generated by binary black holes, each of 2.5 solar masses. Signals  B3 and B4 are produced by  
binary black holes with mass ratio 3:1. The signal B3 does not include spin effects while B4 includes rapid spinning of both 
bodies.

The minimum possible values of $\crm$ are given in 
 the eighth column of Table~\ref{TABLE6}. In this case 
$\gqm > \lqm$ for all the signals. Hence, the 
DC was set to achieve $\gqm=\gqd$  
(second and third columns of Table~\ref{TABLE6}).  
These values produced $\lqd > \lqm$ for all the signals. 
The remarkable performance of the DC 
 emerges from the 
$\crd$ figures listed in the last column 
of Table~\ref{TABLE6}.
\begin{table}[!h]
\begin{center}
\begin{tabular}{|c||r|r|r|r|r|r|r|r||}
\hline
Signal&$\gqm$& $\gqd$&$\lqm$&$ \sdm$& $\lqd$&$\sdd$&$\crm$&$ \crd$ \\ \hline
B1 & 70.2 & 70.2 & 66.2&  8.5 & 67.2 &8.0& 2.7 & 51.7\\ \hline
B2 & 70.3 & 70.3& 65.2  & 10.4 & 66.5  &9.5& 2.6 & 34.2\\ \hline
B3 & 59.0 & 59.0& 58.5 & 2.0 &58.8 & 0.6 & 2.7 & 29.3\\ \hline
B4 & 55.6 & 55.6&54.9 & 2.1 & 55.2 & 0.7 & 2.7 & 28.9 \\ \hline  \hline
\end{tabular}
\caption{Comparison of the $\crm$ and $\crd$ values
 for the audible part of the signals in the Binary group. 
  The second column gives the values of
 $\gqm$ produced by the MP3 signal recovered from files
with
$\crm$  as listed the 8th column.
The  3rd column are the identical values of
 $\gqd$  produced
 by the signal recovered from the DC files with
 $\crd$ as listed in
 the last column.  The 4th and 6th columns
 are the values of $\lqm$ and $\lqd$, respectively. The
5th and 7th columns are the corresponding standard deviations.}
\label{TABLE6}
\end{center}
\end{table}

\begin{table}[!h]
\begin{center}
\begin{tabular}{|c||r|r|r|r|r|r|r|r||}
\hline
Signal&$\gqm$& $\gqd$&$\lqm$&$ \sdm$& $\lqd$&$\sdd$&$\crm$&$ \crd$ \\ \hline
B1 & 68.8 & 68.8 & 65.9&  8.6 & 65.9 &7.8& 5.4 & 56.2\\ \hline
B2 & 68.6 & 68.6& 65.0  & 10.5 & 65.0  &9.4& 5.2 & 36.9\\ \hline
B3 & 53.4 & 54.1& 53.9 & 2.9 &53.9 & 0.4 & 5.5 & 37.5\\ \hline
B4 & 49.3 & 49.6&49.7 & 3.4 & 49.7 & 0.4 & 5.5 & 39.3 \\ \hline  \hline
\end{tabular}
\caption{Same description as Table~\ref{TABLE6} but 
 for double values of $\crm$.}
\label{TABLE7}
\end{center}
\end{table}
\section{Discussion}
The results of Tables~\ref{TABLE1}-\ref{TABLE7} demonstrate  the  remarkable compression power of the DC,  in comparison 
 to MP3  at equivalent 
high quality of the recovered signal. For all 
cases the relation
\be
\label{gam}
\crd= \gamma \,\,\crm
\ee
holds for values of $\gamma$ varying from a minimum 
value  $\gamma_{\min}=1.7$ for signal S12  in 
Table~\ref{TABLE5} to 
a maximum value  $\gamma_{\max}= 19.2$ for signal B1 in 
Table~\ref{TABLE6}.  
Table~\ref{TABLE8} shows the mean value $\bar \gamma$ 
 in each of the above 
tables. It also shows  the corresponding 
$\std$ as well as the values of 
 $\gamma_{\min}$ and $\gamma_{\max}$
in each of the tables.

\begin{table}[!h]
\begin{center}
\begin{tabular}{|c||r|r|r|r||}
\hline
Table& $\bar \gamma$&$\std$& $\gamma_{\min}$& $\gamma_{\max}$ \\ \hline
1 & 7.7 &2.3 & 5.9 &11.0\\ \hline
2 & 2.8 & 0.3& 2.4 & 3.4 \\ \hline
3 & 2.1  & 0.2 & 1.8 & 2.4 \\ \hline
4 & 2.9  & 0.7& 2.2  & 5.1 \\ \hline
5 &  2.2 & 0.6  & 1.7 & 3.2\\ \hline  
6 & 13.5 & 3.9 & 10.8 & 19.2 \\ \hline 
7 & 7.7 & 1.4 & 6.8 & 9.7\\ \hline  \hline
\end{tabular}
\caption{Statistic of the factor $\gamma$ (c.f. 
\eqref{gam}) for Table~\ref{TABLE1}-\ref{TABLE7}. 
The second column corresponds to the mean value 
with respect to the signals in each table. 
The third column gives the corresponding standard deviation.}
\label{TABLE8}
\end{center}
\end{table}

It is worth commenting that the component 
of the dictionary containing the atoms of 
small support plays an essential role of achieving 
large values of $\crd$ for high quality recovery. 
This is much more important for the   
group of EMRI sound. For example, if the compression of 
signal E1 in Table~\ref{TABLE2} is carried out 
in the same way but excluding the sub-dictionary
with those atoms, the value of $\crd$ drops from 
 6.8 to 3.5. Nevertheless, even if for approximating E1 
at the quality of Table~\ref{TABLE2} the percentage of 
atoms of small support is significant ($41\%$) 
the contribution to the signal approximation is  
minor. The norm of the signal E1 is 
  207.6 and the norm of component generated by the atoms of 
small support is only 2.1. This  component of small norm is 
needed to achieve the high values of $\lqd$ and $\gqd$
required in this study. 
Contrarily, for compression of lower 
 quality recovery these atoms  play no role for 
 most signals.  
 At lower quality, however, the power of the 
DC increases substantially. For example, 
when compressing with MP3 and $\crm=8$ the signal E1 
the recovered signal produces $\lqm=41.3$ dB. The
DC achieves the same value of $\lqd$ (and 
$\gqd>\gqm$) for $\crd=53!$ 

\subsection{Beyond Compression}
\begin{figure}[!ht]
\begin{center}
\includegraphics[width=8cm,height=5.5cm]{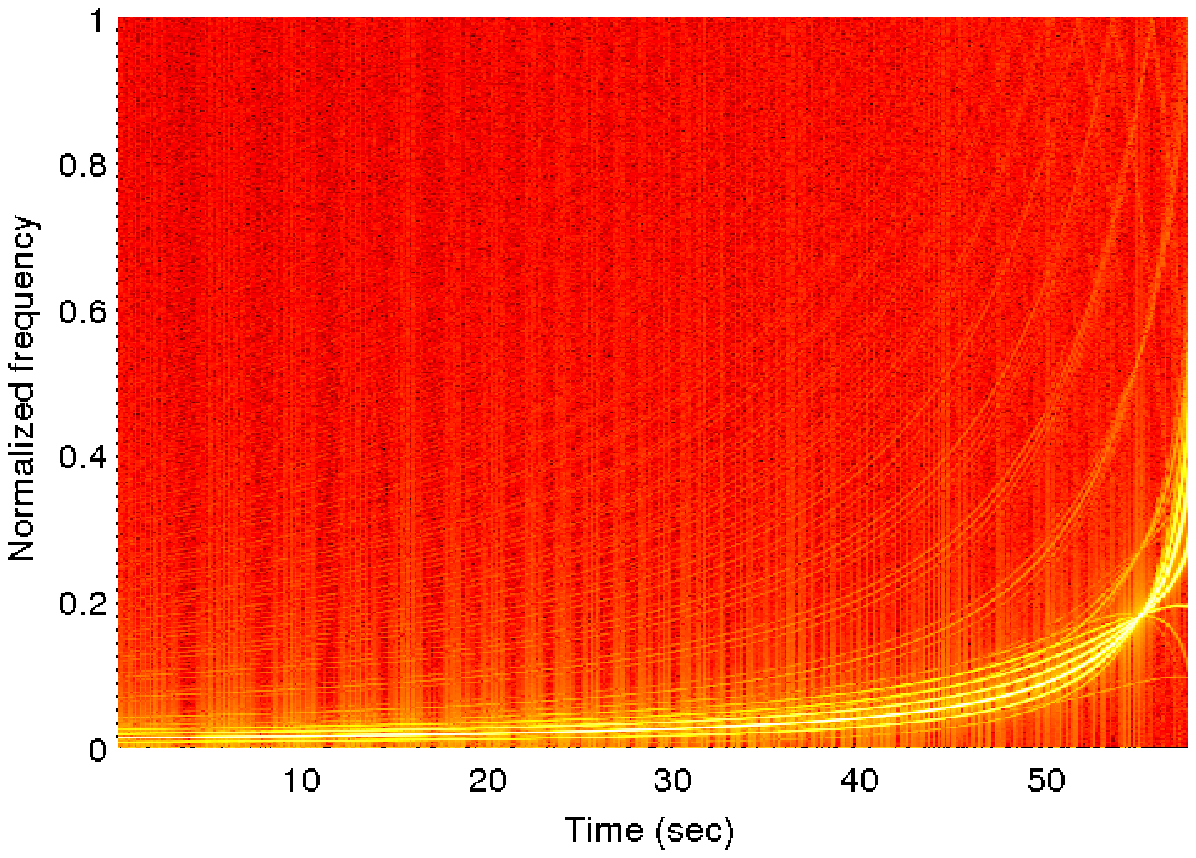}
\includegraphics[width=8cm,height=5.5cm]{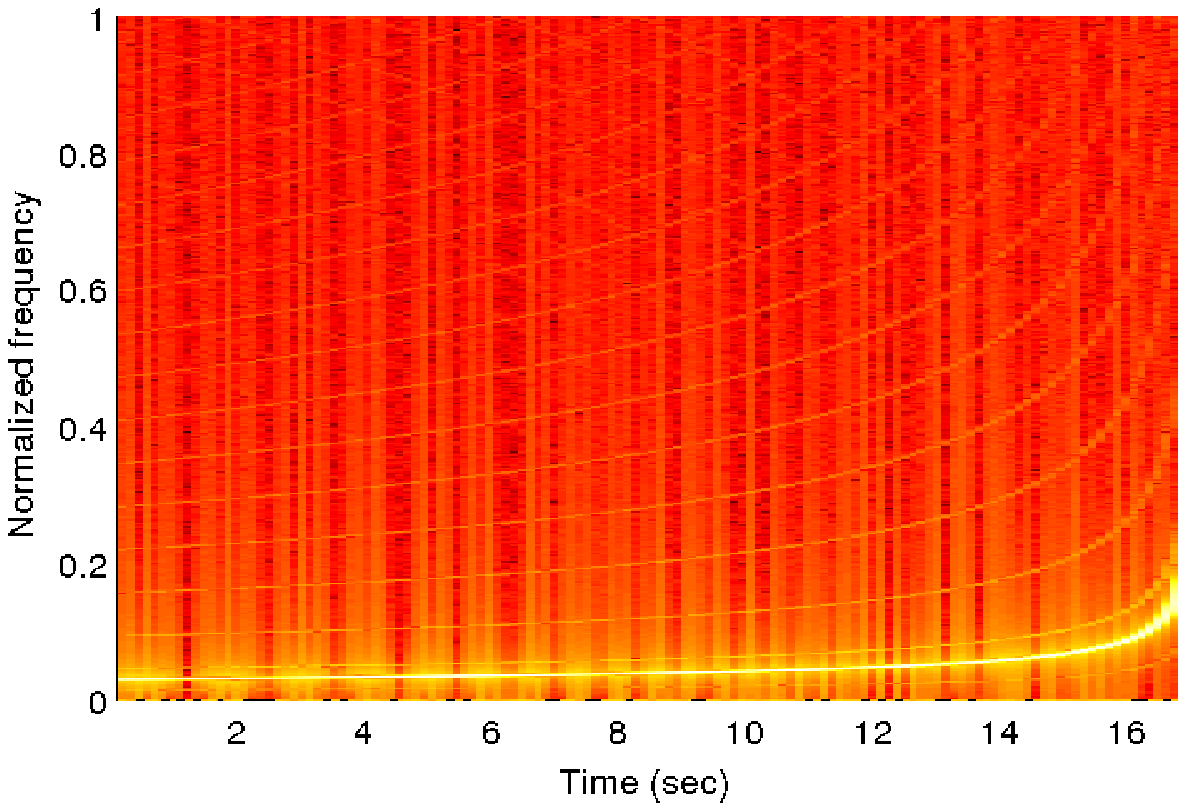}\\
\includegraphics[width=8cm,height=5.2cm]{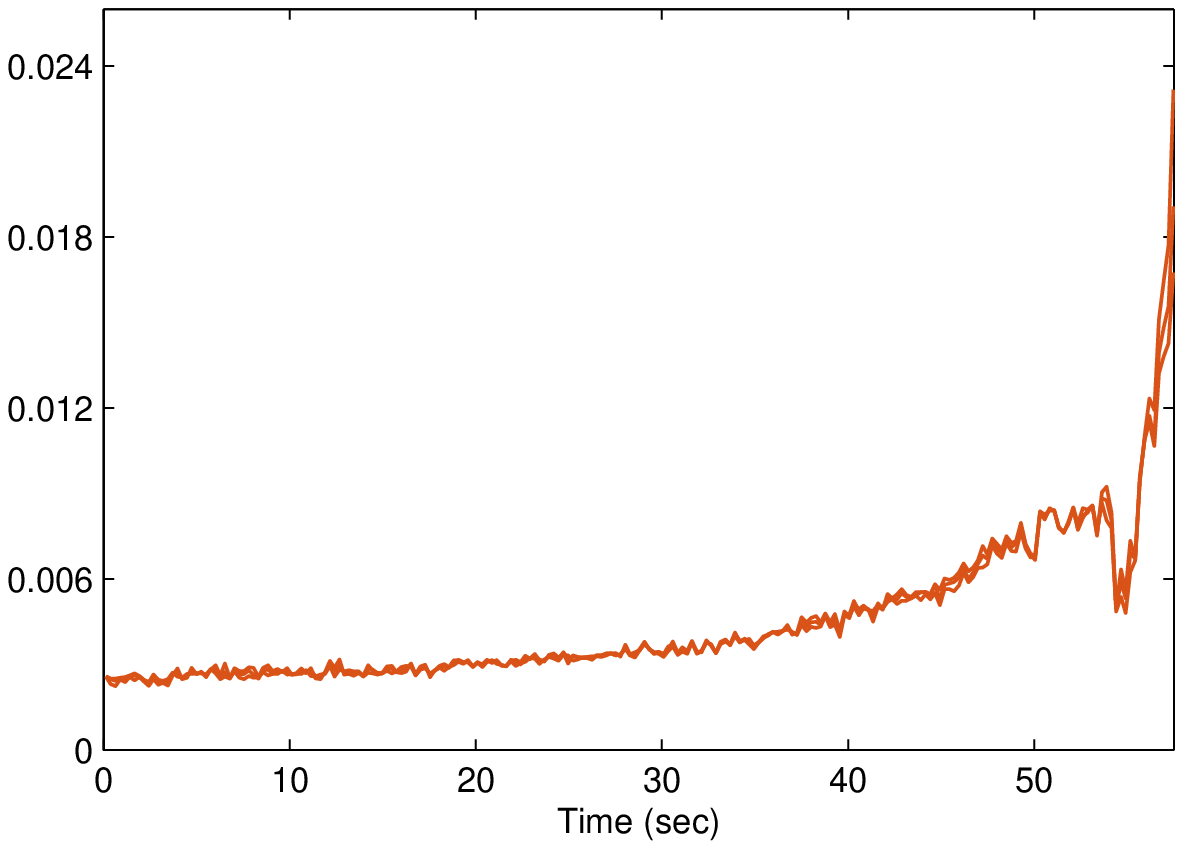}
\includegraphics[width=8cm,height=5.2cm]{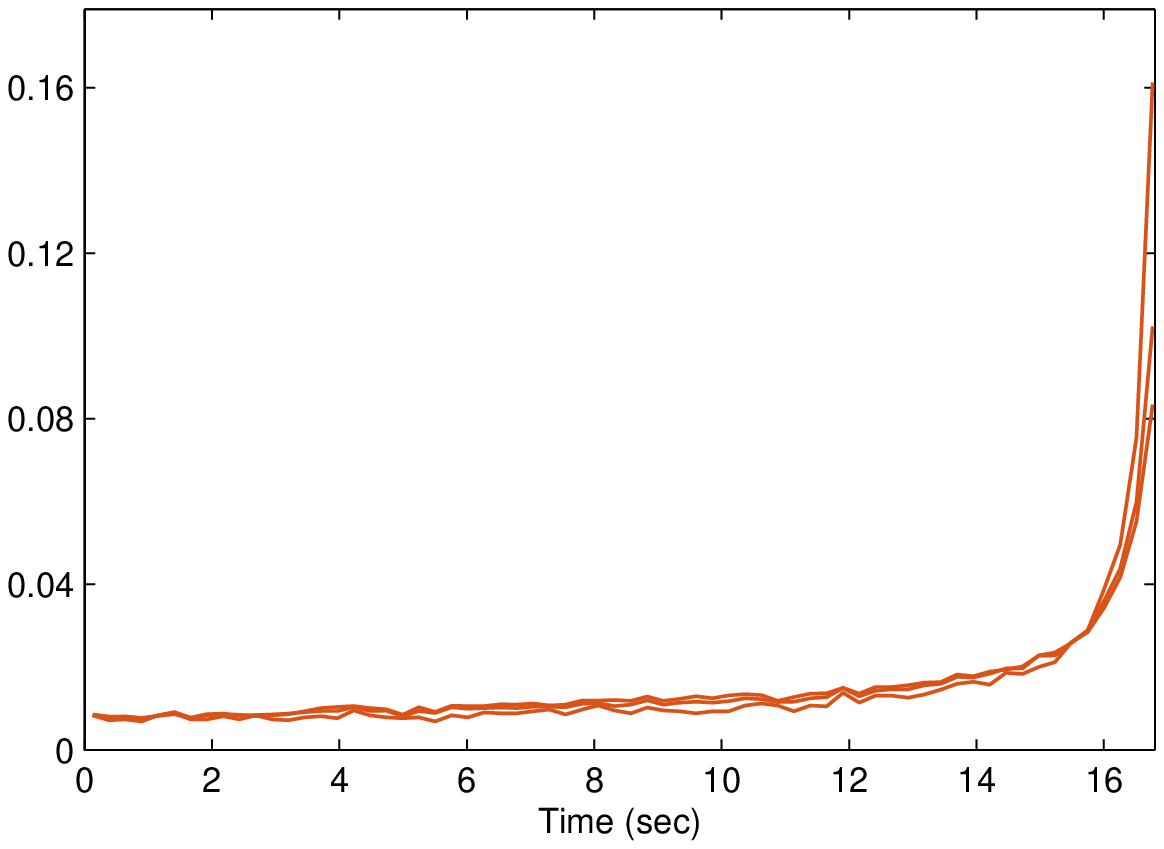}
\caption{\small{
The upper graphs are the spectrograms of the 
signals E1 (left) and S1 (right).
The lines in the lower graphs 
join the values $\kqt=1,\ldots, Q$ resulting
when approximating these signals at
three different qualities: $\lqd= 65$, 60, and 55 dB. 
In the left graph each of the three lines 
joins $Q=225$ points. In the right graph $Q=66$ points.}}
\label{fing}
\end{center}
\end{figure}

The success of the DC in producing a small file stems from 
 the ability of constructing a signal approximation of 
good quality, but 
involving less elementary components than the 
number of samples giving the signal. Let's 
suppose that to represent at the desired quality  
the block $\vf_q$ in the signal partition one needs 
$\kq$ dictionary atoms, and let's normalize these values 
$\kqt= {\kq}/\sum_{q=1}^Q {\kq}$ for 
comparison purposes. As shown in Fig.~\ref{fing} 
the numbers $\kqt,\,q=1,\ldots,Q$
 render meaningful information about the signal internal
 variations over time. In the lower graphs of 
Fig.~\ref{fing} each of these values is 
  located in the horizontal axis at
 the center of the corresponding block (of size $\Nq=2048$)
 and provides a condensed digital summary of the sound. 
The  left upper graph is 
the spectrograms of the signal E1. The lines in 
the lower left graph 
 join the values $\kqt,\, q=1,\ldots, 225$ resulting 
when approximating this signal at
three different qualities: $\lqd= 65$, 60 and 55 dB. As
 observed in the graph, the three lines are very close 
to each other and 
 account for the node in the spectrogram  which occurs 
 at around 55 secs. The  right upper graph of Fig.\ref{fing}
is the spectrogram of the signal S1. This signal differs 
from the signal E1 only in the spin ($35.94\%$ of the maximum value). Also in this case, the three lines connecting  
 the values $\kqt,\, q=1,\ldots, 66$ for 
approximations of $\lqd= 65$, 60, and 55 dB  are 
close to each other and 
 account for the main feature of the 
spectrogram, which is characterized by a rise of frequency 
towards the end.  
It is an interesting feature of 
the digital summary, given by the points in the lower 
graphs, the small cardinality in 
relation to the length of the signal: 
225 points for the signal E1 
given by $N=460800$ samples and 66 points for the 
signal S1 given by $N=135168$ samples. 
\section{Conclusions}
A dedicated codec for compression of gravitational sound 
with high quality recovery has been proposed. 
The power of the format to store this type of sound 
stems from the mathematical model
 for representing the signal. The model 
is based on recursive selection 
of elementary components approximating the signal 
 with accuracy. 
The components, called atoms, are chosen 
from a redundant set, called a dictionary, which is 
the union of two sub-dictionaries consisting 
of atoms of different nature. One sub-dictionary contains 
trigonometric atoms. The other sub-dictionary contains 
pulses of small support. While the contribution 
of the atoms of small support to the signal approximation  is minor, they are necessary to obtain approximations of 
 high accuracy. 
At lower quality recovery (less than 50 dB)  
 this sub-dictionary
 could be avoided and the compression power of the proposed 
 codec would be even more potent. Nonetheless, the study 
reported here 
focuses on compression with high quality point-wise 
 recovery. 

The proposal was tested on the sound representation 
of the  detected short chirp {\tt{gw151226}},  
 and on a set of longer
signals numerically simulated at MIT. 
Comparisons with the compression standard MP3 
resulted in a significant increment of compression 
power for equivalent quality of the recovered 
signal. As a byproduct, 
the codec generates a condensed digital summary of the 
sound which could be of assistance for 
identification tasks. 

{\bf{Note:}} 
All the results in this paper can be reproduced using the
MATLAB software which has been made available  on
\cite{webpage}.

\section*{Acknowledgment}
Thanks are due to Karl Skretting for making available the
Arith06 function \cite{Karl} for
arithmetic coding, which has been used
 at the entropy coding step. The author is 
grateful to Prof Hughes and the people from
his group who participated in the
simulation of the signals used in this study.


\end{document}